\documentclass[11pt]{article}
\usepackage{fullpage,hyperref,amssymb,amsmath,url,graphicx}
 
\def\Bhat{\mathrm{Bhat}}
\def\calE{\mathcal{E}}
\def\calX{\mathcal{X}}
\def\bbN{\mathbb{N}}
\def\KL{\mathrm{KL}}
\def\dx{\mathrm{d}x}
\def\ds{\mathrm{d}s}
\def\floor#1{\left\lfloor {#1}\right\rfloor}
\def\calP{\mathcal{P}}

\newtheorem{remark}{Remark}
\newtheorem{Example}{Example}
\newtheorem{Theorem}{Theorem}
\newtheorem{Corollary}{Corollary}
\newtheorem{Proposition}{Proposition}

\title{A note on some information-theoretic divergences\\ between Zeta distributions}

\author{Frank Nielsen\\ Sony Computer Science Laboratories Inc.\\
Tokyo, Japan}

\date{}

\begin{document}
\maketitle

\begin{abstract}
We consider the zeta distributions which are discrete power law distributions that can be interpreted as the counterparts of the continuous Pareto distributions with unit scale.
The family of zeta distributions forms a discrete exponential family with normalizing constants expressed using the Riemann zeta function. We report several information-theoretic measures between zeta distributions and study their underlying information geometry. 
\end{abstract}

\section{Introduction}

The zeta distributions~\cite{Encyclopedia-2005,goldstein2004problems} are parametric discrete distributions with probability mass functions (PMFs) defined on the support of the natural integers $\bbN$ indexed by a scalar  parameter $s\in (1,\infty)$ as follows:
$$
p_s(x)=\Pr[X=x]\propto\frac{1}{x^s},\quad x\in\calX=\bbN=\{1,2,\ldots\}.
$$ 
The normalization function $\zeta(s)$ of the zeta distributions $p_s(x)=\frac{1}{\zeta(s)}\, \frac{1}{x^s}$ is the real Riemann zeta function~\cite{edwards1974riemannaes,titchmarsh1986theory,patterson1995introduction,zeta-2014}:
$$
\zeta(s)=\sum_{i=1}^\infty \frac{1}{i^s}, \quad s>1.
$$
\begin{figure}[b]
\centering
\includegraphics[width=0.45\textwidth]{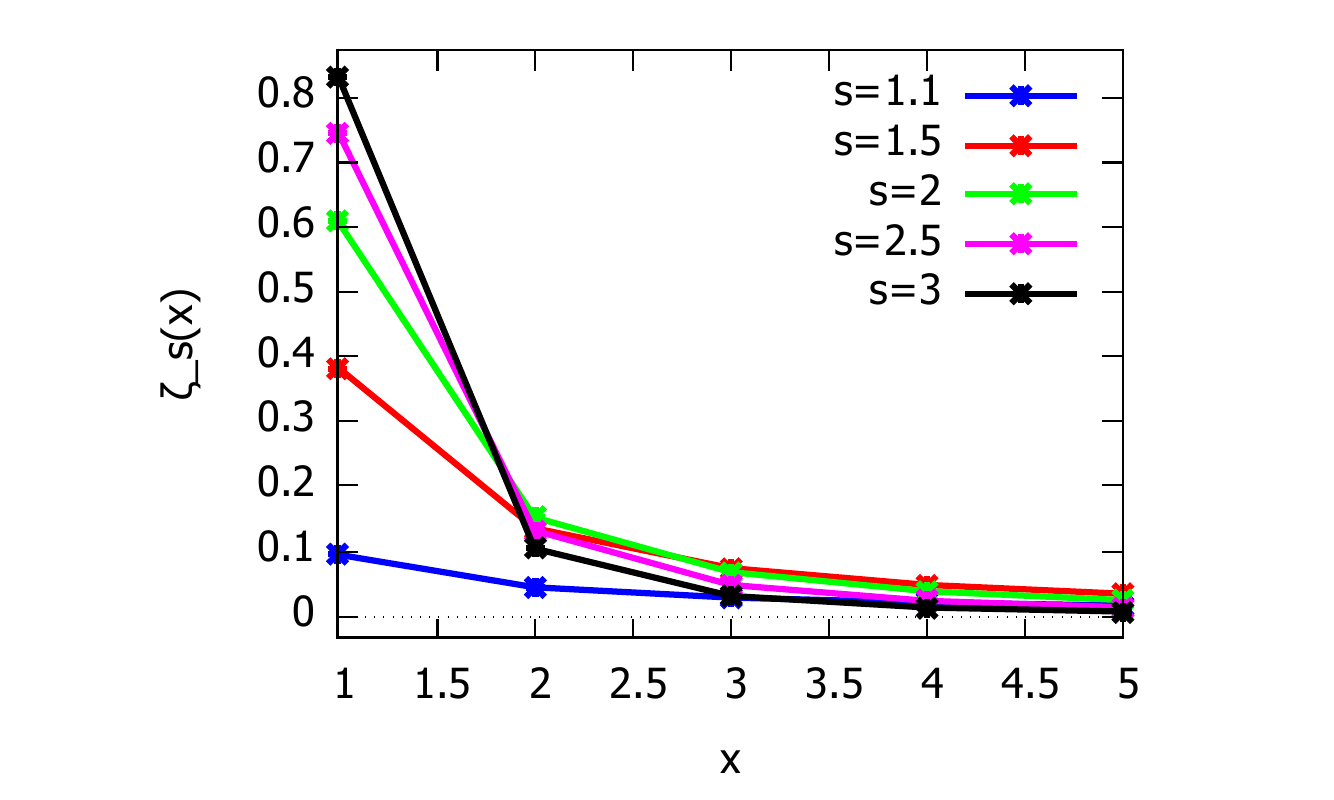}
\caption{Several plots of the probability mass functions of zeta distributions.}\label{fig:plotzetadist}
\end{figure}

\begin{figure} 
\centering
\begin{tabular}{cc}
\includegraphics[width=0.45\textwidth]{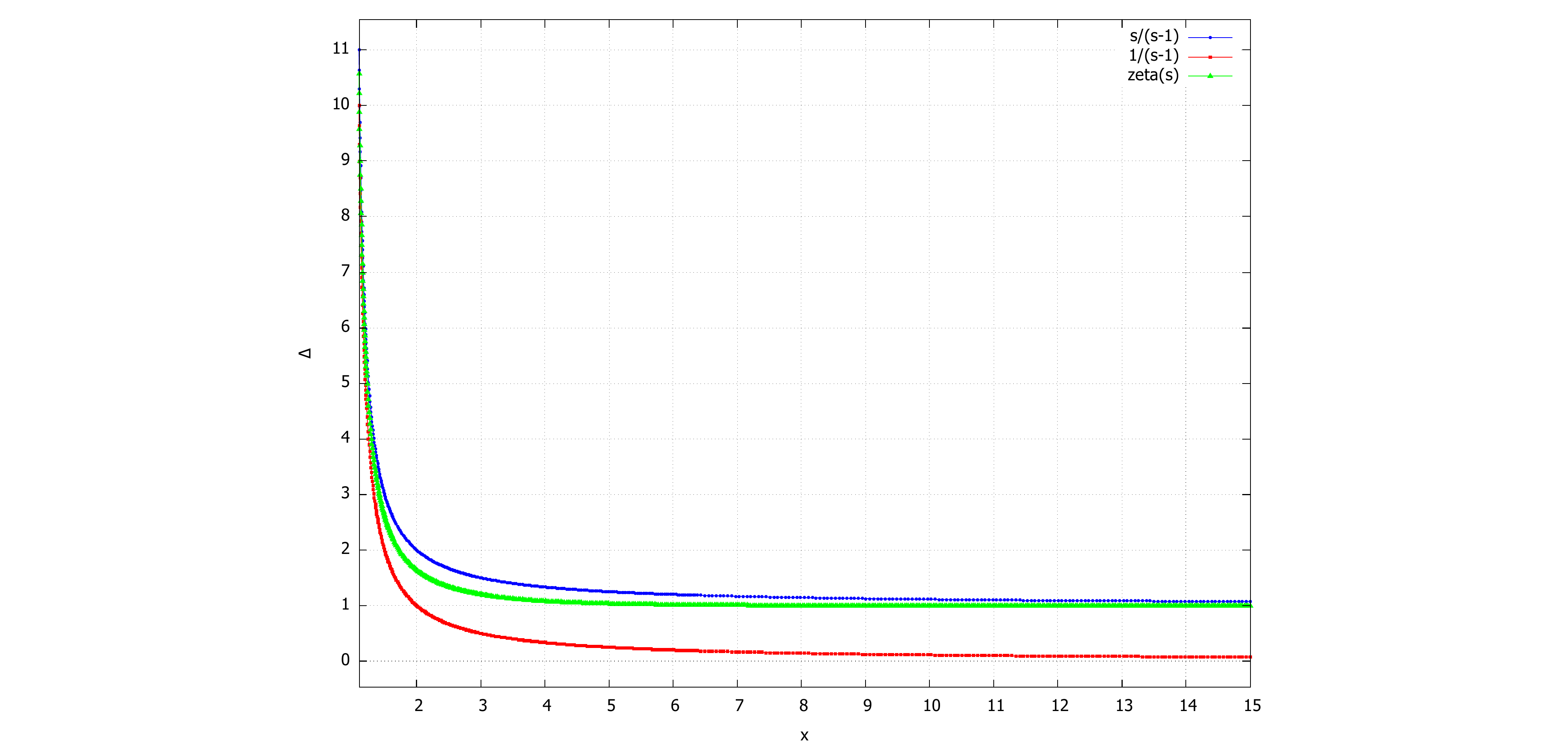} &
\includegraphics[width=0.45\textwidth]{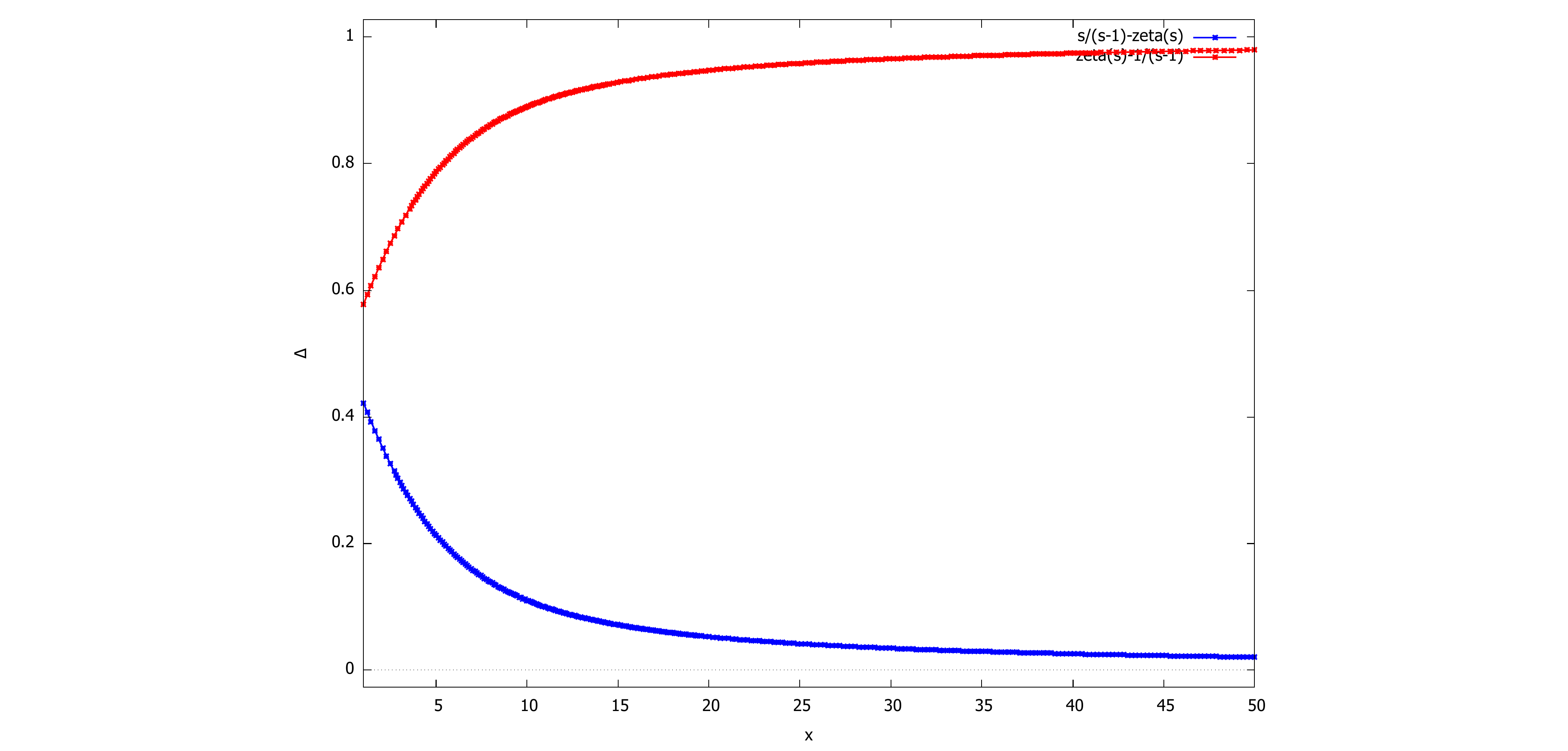}\\
(a) & (b)
\end{tabular}
\caption{(a) Plot of the zeta function $\zeta(s)$ with its lower and upper bounds of Eq.~\ref{eq:ineqzeta}. (b) Plots of the lower and upper bound gaps.}\label{fig:plotlubounds}
\end{figure}

Figure~\ref{fig:plotzetadist} displays several PMFs of zeta distributions.
The zeta function can be bounded as follows~\cite{rfi2008power} (see Figure~\ref{fig:plotlubounds}):
\begin{equation}\label{eq:ineqzeta}
\frac{1}{s-1}\leq \zeta(s)\leq \frac{s}{s-1}.
\end{equation}

The set of zeta distributions $\mathcal{Z}=\{p_s(x)\ :\ s\in(1,\infty)\}$ forms a {\em discrete exponential family}~\cite{EF-2014} with natural parameter $\theta(s)=s$ lying in the natural parameter space $\Theta=(1,\infty)$, 
 the sufficient statistic $t(x)=-\log x$, and the cumulant function or log-normalizer $F(\theta)=\log\zeta(\theta)$, a strictly convex and real analytic function\footnote{The zeta function on the complex plane is a meromorphic function  with a
simple pole at $s=1$.} (and hence $\zeta(s)$ is log-convex).
Thus the pmf of zeta distributions can be rewritten in the canonical form of exponential families as:
$$
p_s(x)=\exp\left(\theta(s)t(x)-F(\theta(s))\right).
$$
The characteristic function is thus $\phi_s(t)=\frac{\zeta(s+it)}{\zeta(s)}$.
An acceptance/rejection method to sample zeta variates is presented in Appendix~\ref{sec:variate}.
Thus as an exponential family, the zeta distributions are maximum entropy discrete distributions for the constraint $-E[\log x]=\eta$ (a result formerly derived in~\cite{guiasu1986optimization}):

$$
\max_p \left\{H(p)\quad :\quad -E_p[\log x]=\eta \Rightarrow p=\mathrm{Zeta}(\theta(\eta))\right\},
$$
where $H(p)$ denotes the Shannon entropy of any distribution with full support $\bbN$:
$$
H(p)=-\sum_{x\in\bbN} p(x)\log p(x).
$$
We get the dual moment parameterization of a zeta distribution:
$$
\eta(\theta)=F'(\theta)=\frac{\zeta'(\theta)}{\zeta(\theta)}.
$$

A zeta distribution $p_s(x)$ can be interpreted as the discrete equivalent of a Pareto distribution $q_s(x)$ of scale $1$ and shape $s-1$ 
with probability density function $q_s(x)=\frac{s-1}{x^s}$ for $x>1$.
See Table~\ref{tab:comparison}.

\begin{table}
\centering
\caption{Comparisons between the Zeta family and the Pareto subfamily. The function $\zeta(s)$ is the real zeta function.}\label{tab:comparison}
\vskip 0.5cm
{\footnotesize
\begin{tabular}{|l|l|l|}\hline
 & Zeta distribution & Pareto distribution\\ \hline
\multicolumn{3}{|c|}{Exponential family $\exp(\theta t(x)-F(\theta))$}\\ \hline
& Discrete EF & Continuous EF\\ \hline
PMF/PDF & $p_s(x)=\frac{1}{x^s\zeta(s)},\quad \zeta(s)=\sum_{i=1}^\infty \frac{1}{i^s}$ & $q_s(x)=\frac{s-1}{x^{s}}$\\
Support $\calX$ & $\bbN=\{1, 2, \ldots\}$ & $(1,\infty)$\\
Natural parameter $\theta$ & $s\in\Theta=(1,\infty)$ & $s\in\Theta=(0,\infty)$\\
Cumulant $F(\theta)$ & $\log\zeta(\theta)$  & $-\log(\theta-1)$\\
Sufficient statistic $t(x)$ & $-\log x$ &   $-\log x$\\ \hline
Moment parameter $\eta=-E[\log x]$ & $\frac{\zeta'(\theta)}{\zeta(\theta)}$ & $-\frac{1}{s-1}$ \\
Mean &  $\frac{\zeta(s-1)}{\zeta(s)}$ & $\frac{s}{s-1}$\\
Variance & $\frac{\zeta(s)\zeta(s-2)-\zeta(s-1)^2}{\zeta(x)^2}, s>3$ & $\frac{s}{(s-1)^2(s-2)}, s>2$ \\
Conjugate $F^*(\eta)$ &  $-H[p_s]=-\sum_{i=1}^\infty  \frac{1}{i^s\zeta(s)}\log(i^s\zeta(s))$  & $\eta-1-\log(-\eta)$\\ \hline
Maximum likelihood estimator & $\hat{\eta}=\frac{\zeta'(\hat{\theta})}{\zeta(\hat{\theta})}=-\frac{1}{n}\sum_{i=1}^n \log x_i$ & $\hat{s}=\frac{n}{\sum_{i=1}^n \log x_i}$\\ \hline
Fisher information & $\sum_{i=0}^\infty \Lambda(i)\log(i)i^{-s}$ & $\frac{1}{(s-1)^2}$ \\
Entropy $-F^*(\eta(s))$ & $\sum_{i=1}^\infty  \frac{1}{i^s\zeta(s)}\log(i^s\zeta(s))$  & $1+\frac{1}{s-1}-\log(s-1)$\\
Bhattacharyya coefficient $I_\alpha$ & $\frac{\zeta(\alpha s_1+(1-\alpha)s_2)}{\zeta(s_1)^\alpha \zeta(s_2)^{1-\alpha}}$ &
$\frac{\alpha s_1+(1-\alpha)s_2}{s_1^\alpha s_2^{1-\alpha}}$ \\
Kullback-Leibler divergence $D_\KL$ & $\log(\zeta(s_2)) -  \sum_{i=1}^\infty  \frac{1}{i^s\zeta(s)}\log(i^s\zeta(s)) - s_2 \frac{\zeta'(s_1)}{\zeta(s_1)}$ & $\log\left(\frac{s_1-1}{s_2-1}\right)+\frac{s_2-s_1}{s_1-1}$\\ \hline
\end{tabular}
}
\end{table}

The zeta function can be calculated fast~\cite{borwein2000efficient,fastZeta-2011} and precisely~\cite{cohen1992calcul,preciseZeta-2015}.
The derivatives of the zeta function have also been studied~\cite{yildirim1996note,fastZeta-2011}.
In particular, for all even positive integer values, the zeta function $\zeta(2n)$ can be evaluated {\em exactly} using Bernoulli numbers $B_{2n}$~\cite{ConcreteMath-1994} (\S 6.5, p. 283):
$\zeta(2n)=\frac{(-1)^{n+1} B_{2n}(2 \pi)^{2 n}}{2(2 n) !},\quad n\in\bbN$. 
For example,  we have $\zeta(2)=\frac{\pi^2}{6}$, $\zeta(4)=\frac{\pi^4}{90}$, 
$\zeta(6)=\frac{\pi^6}{945}$, etc. 

The zeta distributions are related to the Zipf distributions~\cite{powers1998applications,saichev2009theory}
$p_{s,N}(x)\propto \frac{1}{x^s}$ for $x\in\{1,\ldots,N\}$
 and the Zipf-Mandelbrot distributions~\cite{mandelbrot1961theory,mandelbrot1966information,ZipfMandelbrotFDiv-2018} 
$p_{s,q,N}(x)\propto \frac{1}{(x+q)^s}$ for $x\in\{1,\ldots,N\}$
which play an important role in quantitative linguistics.
The Zipf distributions and the Zipf-Mandelbrot distributions both have finite support and can be interpreted as truncated zeta distributions (right truncation for Zipf distributions and both left \& right truncations for the Zipf-Mandelbrot distributions) with normalizing constants which can be calculated approximately using properties of the zeta function~\cite{naldi2015approximation}.
Left-only truncations of the Zeta distributions are called Hurwitz zeta distributions~\cite{hu2006hurwitz}.
Similarly, truncated Pareto distributions are used in applications~\cite{deluca2013fitting}.
Table~\ref{tab:trunczeta} summarizes the terminology of truncated zeta distributions.
Notice that truncated distributions of an exponential family with fixed truncation support form another 
exponential family~\cite{duodivergence-2022}. 
Notice that the natural parameter space $\Theta$ of Zipf distributions is $(0,\infty)$ while the natural parameter space of zeta distributions is $(1,\infty)$ due to convergence requirements for the infinite zeta summation. 

\begin{table}
\centering
\caption{Terminology for the truncated zeta distributions.}\label{tab:trunczeta}
\vskip 0.5cm
\begin{tabular}{lll}
Left truncation & right truncation & distribution name \\ \hline
Yes & Yes & Zipf-Mandelbrot distribution~\cite{ZipfMandelbrotFDiv-2018}\\
Yes & No & Hurwitz zeta distribution~\cite{hu2006hurwitz}\\
No & Yes & Zipf distribution~\cite{saichev2009theory}
\end{tabular}
\end{table}

The zeta distributions and its related Hurwitz/Zipf/Zipf-Mandelbrot/distributions are discrete power law distributions which can be used to model the frequency of a word as a power law function of its frequency rank~\cite{baayen2001word}.
For example, the rank-frequency datasets of the translations of the Holy Bible in $100$ languages have been analyzed using the Zipf distributions in~\cite{ZipfHolyBible-2017}.
Zipf's law occur empirically in many datasets where the ranked data exhibit the higher the fewer property (e.g., US firm sizes~\cite{axtell2001zipf} or the surname frequencies~\cite{SurnameZipf-1983}).
The Zipf's law is empirically only an approximation of a more complex distribution (see~\cite{moreno2016large} for a study on the 30k english texts of the Project Gutenberg).

The zeta distributions are infinite divisible~\cite{hu2006hurwitz,saito2012note,devianto2019characterization,ZetaDistribution-2019}: A random variable following a zeta distribution can be expressed as the probability distribution of the sum of an arbitrary number of independent and identically distributed random variables. 
In applications, it is important to quantitatively discriminate between zeta distributions (see, for example~\cite{wang2013near,oosawa2014sql} or~\cite{doray1995quadratic}). Mixtures of zeta distributions have also been used to model social networks~\cite{jung2021mixture}.
In general, products of exponential families yield other exponential families. The products of $d$ zeta distributions form an exponential family called the Shintani multidimensional zeta distributions~\cite{aoyama2013multidimensional} or the zeta-star distributions~\cite{saito2012note}.

We study information-theoretic divergences between zeta distributions by considering the fact that the set of zeta distributions form a discrete exponential family~\cite{barndorff2014information}.

\section{Amari's $\alpha$-divergences and Sharma-Mittal divergences}

To analyze sets of datasets exhibiting power law distributions, we may consider a notion of dissimilarity between discrete power law distributions. For example, a $100$ language translations of the Holy Bible was considered in~\cite{ZipfHolyBible-2017} where each translation was analyzed  by a Zipf distribution of the word rank-frequencies and characterized by the Zipf power exponent $s$ (see Table~1 in~\cite{ZipfHolyBible-2017}). In order to cluster hierarchically or by $k$-means this set of (approximate) zeta distributions, we need to define a notion of distance between zeta distributions. 

\begin{figure}
\centering
\includegraphics[width=0.75\textwidth]{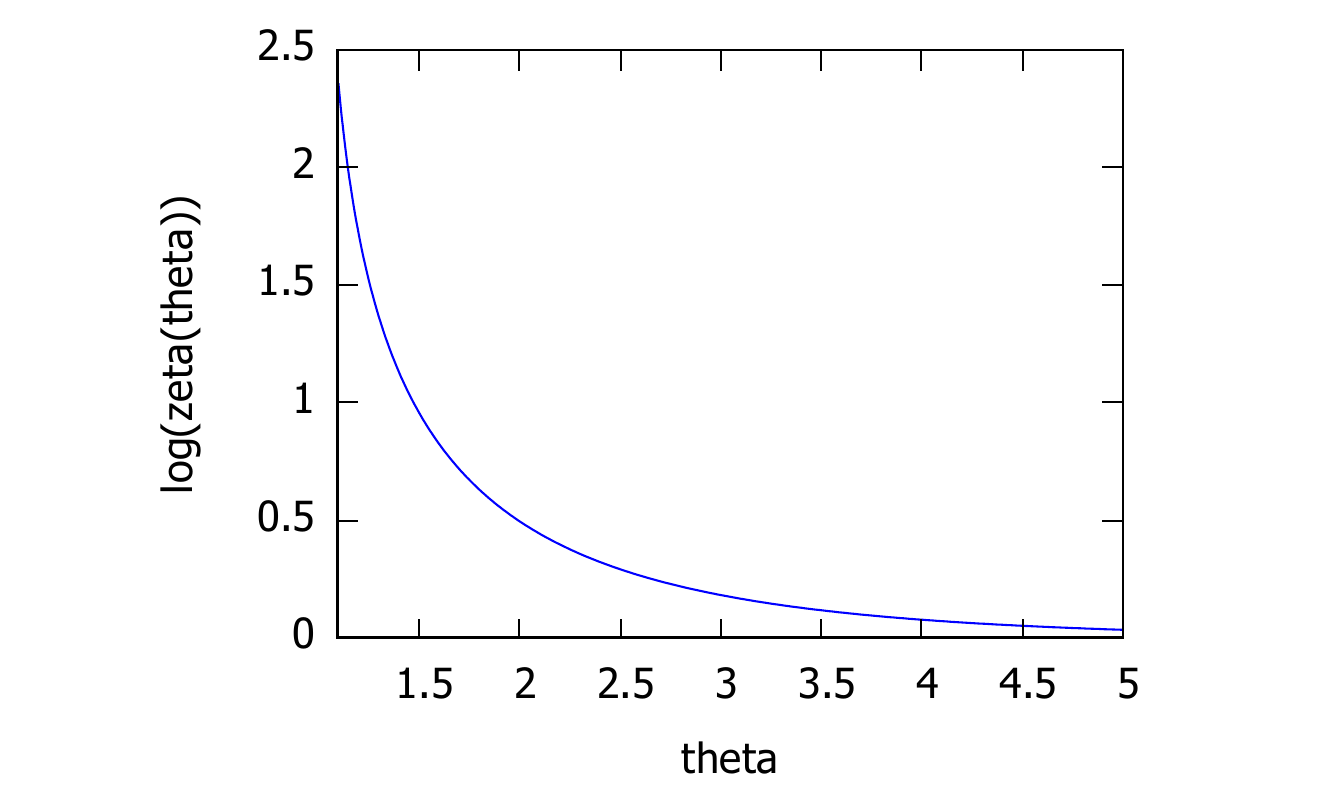}
\caption{Plot of $F(\theta)=\log\zeta(\theta)$, a strictly convex and analytic function.}\label{fig:plotF}
\end{figure}

To measure the dissimilarity between two zeta distributions $p_{s_1}$ and $p_{s_2}$, 
one can use the {\em $\alpha$-divergences}~\cite{alphadiv-2010} defined for a real $\alpha\in (0,1)$ as follows:
$$
D_\alpha[p_{s_1}:p_{s_2}] := \frac{1}{\alpha(1-\alpha)}\left(1-I_\alpha[p_{s_1}:p_{s_2}]\right),
$$
where
$$
I_\alpha[p_1,p_2]:= \sum_{i=1}^\infty p_{1}(x)^\alpha p_{2}(x)^{1-\alpha},\quad \alpha\in (0,1)
$$
is the $\alpha$-Bhattacharyya coefficient.
The set of zeta distributions $\mathcal{E}:=\{p_s(x)\ :\ s\in(1,\infty)\}$ form a {\em discrete exponential family}~\cite{barndorff2014information} with natural parameter $\theta(s)=s$ (natural parameter space $\Theta=(1,\infty$), 
sufficient statistic $t(x)=-\log x$, and cumulant function $F(\theta)=\log\zeta(\theta)$ (see Table~\ref{tab:comparison}), a strictly convex and analytic function (see Figure~\ref{fig:plotF}):
$p_s(x)=\exp\left(\theta(s)t(x)-F(\theta(s))\right)$.

It follows from~\cite{BR-2011} that the skewed Bhattacharyya coefficient amounts to a skewed Jensen divergence between the natural parameters of the exponential family $\mathcal{E}$:
$$
I_\alpha[p_{s_1}:p_{s_2}] = \exp\left(-J_{F,\alpha}(s_1:s_2)\right),
$$
where $J_{F,\alpha}$ is the skewed Jensen divergence induced by a strictly convex and smooth convex function $F(\theta)$:
\begin{eqnarray*}
J_{F,\alpha}(s_1:s_2)&:=&\alpha F(s_1)+(1-\alpha)F(s_2)-F(\alpha s_1+(1-\alpha)s_2)\geq 0,\\
&=& \log\left( \frac{\zeta(s_1)^\alpha \zeta(s_2)^{1-\alpha}}{\zeta(\alpha s_1+(1-\alpha)s_2)}\right).
\end{eqnarray*}

Thus we have the $\alpha$-divergences between two zeta distributions $p_{s_1}$ and $p_{s_2}$ available in closed-form.

\begin{Theorem}[$\alpha$-divergences between two zeta distributions]
The $\alpha$-divergence for $\alpha\in (0,1)$ between two zeta distributions $p_{s_1}$ and $p_{s_2}$ is:
$$
D_\alpha[p_{s_1}:p_{s_2}] = \frac{1}{\alpha(1-\alpha)} \left(1-\frac{\zeta(\alpha s_1+(1-\alpha)s_2)}{\zeta(s_1)^\alpha \zeta(s_2)^{1-\alpha}}\right).
$$
\end{Theorem}

It follows that when $s_1$, $s_2$, and $\alpha s_1+(1-\alpha)s_2$ are all positive even integers, we can evaluate exactly the $\alpha$-divergences between 
$p_{s_1}$ and $p_{s_2}$.

\begin{Example}
Consider $s_1=4$ and $s_2=12$ with $\alpha=\frac{1}{2}$.
The $\alpha$-divergence for $\alpha=\frac{1}{2}$ is   the squared Hellinger divergence $
D_{\frac{1}{2}}[p_{s_1},p_{s_2}] = \sum_{i=1}^\infty \left(\sqrt{p_{s_1}(i)}-\sqrt{p_{s_2}(i)}\right)^2$.
Since $\alpha s_1+(1-\alpha)s_2=8$, we find the exact squared Hellinger divergence:
$D_{\frac{1}{2}}[p_{4},p_{12}] = 4\left(1-3\sqrt{\frac{715}{6910}}\right)\simeq 0.139929\ldots$.
\end{Example}

Let us report another example where the squared Hellinger divergence is expressed using the zeta function:

\begin{Example}
We consider $s_1=3$, $s_2=7$ and $\alpha=\frac{1}{2}$ so that $\alpha s_1+(1-\alpha)s_2=5$. Then we have
$D_{\frac{1}{2}}[p_{3},p_{7}] = 4\left(1-\frac{\zeta(5)}{\sqrt{\zeta(3)\zeta(7)}}\right)\simeq 0.23261\ldots$
\end{Example}

Since $\lim_{\alpha\rightarrow 1} D_\alpha[p_{s_1}:p_{s_2}]=D_\KL[p_{s_1}:p_{s_2}]$ is the Kullback-Leibler divergence~\cite{alphadiv-2010} (KLD)~\cite{BR-2011}
$$
D_\KL[p_{s_1}:p_{s_2}]:=\sum_{i=1}^\infty p_{s_1}(i)\log\frac{p_{s_1}(i)}{p_{s_2}(i)},
$$
we can approximate the KLD by $D_{1-\epsilon}[s_1:s_2]$ for a small value of $\epsilon$ (say, $\epsilon=1.0e^{-3}$) using fast methods to compute the zeta function~\cite{zetacompute-2011}.
Similarly, when $\alpha\rightarrow -1$, the $\alpha$-divergences tend to the reverse Kullback-Leibler divergence:
$$
\lim_{\alpha\rightarrow -1} D_\alpha[p_{s_1}:p_{s_2}]=D_\KL[p_{s_2}:p_{s_1}]=D_\KL^*[p_{s_1}:p_{s_2}]=\sum_{i=1}^\infty p_{s_2}(i)\log\frac{p_{s_2}(i)}{p_{s_1}(i)}.
$$

\begin{Corollary}[Approximation of the Kullback-Leibler divergence.]\label{cor:klzetaapprox}
The Kullback-Leibler divergence  between two zeta distributions $p_{s_1}$ and $p_{s_2}$ can be approximated for small values $\epsilon>0$ by
$$
D_\KL[p_{s_1}:p_{s_2}]\simeq D_{1-\epsilon}[p_{s_1}:p_{s_2}]= \frac{1}{\epsilon(1-\epsilon)} \left(1-\frac{\zeta((1-\epsilon) s_1+\epsilon s_2)}{\zeta(s_1)^{1-\epsilon} \zeta(s_2)^{\epsilon}}\right).
$$
\end{Corollary}

We can also calculate the KLD $D_\KL[p_{s_1}^{\calX_1}:p_{s_2}^{\calX_2}]$ between two truncated zeta distributions with nested supports 
$\calX_1\subseteq\calX_2$. See~\cite{duodivergence-2022}. A truncated zeta distribution on the support $\{a, a+1,\ldots, b\}\subset\bbN$ (with $b>a$) has pmf $p^{a,b}_s(x)=\frac{p_s(x)}{\Phi_s(b)-\Phi_s(a)}$ where $\Phi_s(u)$ is the cumulative distribution function 
$\Phi_s(u)=\sum_{x\in\{1,\ldots,u\}} p_s(x)=\frac{1}{\zeta(s)} \sum_{x\in\{1,\ldots,u\}} \frac{1}{x^s}$. 

The Chernoff information~\cite{nielsen2013information} is defined by 
$C[p_1,p_2]=-\log \min_{\alpha\in(0,1)} I_\alpha[p_1,p_2]$.
When both pdfs or pmfs belong to the same exponential family, we have~\cite{nielsen2013information}
$$
C[p_{\theta_1},p_{\theta_2}]=J_F(\theta_1:(\theta_1\theta_2)_{\alpha^*})=B_F(\theta_1:(\theta_1\theta_2)_{\alpha^*})
=B_F(\theta_2:(\theta_1\theta_2)_{\alpha^*}),
$$
where $B_F$ denotes the Bregman divergence (corresponding to the KLD) and $(\theta_1\theta_2)_{\alpha^*}=\alpha^*\theta_1+(1-\alpha^*)\theta_2$. For uniorder exponential family like the zeta distributions, we get a closed-form solution~\cite{nielsen2013information}:
$$
\alpha^*=\frac{{F^*}'(\Delta F/\Delta\theta)-\theta_2}{\Delta\theta},
$$
where $\Delta\theta=\theta_1-\theta_2$, $\Delta F=F(\theta_1)-F(\theta_2)$, an $F^*(\eta)$ is the convex conjugate of $F(\theta)$.
For example, applying this formula for the Pareto distributions with $s_1\not=s_2$ and ${F^*}'(\eta)=1-\frac{1}{\eta}$, we get
$$
\alpha^*=\frac{1-s_2}{s_1-s_2}-\frac{1}{\log \frac{s_2-1}{s_1-1}},
$$ 
and the Chernoff information between two Pareto distributions is 
$$
C[p_{s_1},p_{s_2}]=\log \frac{s_1^{\alpha^*}s_2^{1-\alpha^*}}{\alpha^*s_1+(1-\alpha^*)s_2)}.
$$
The information geometry (i.e., the Fisher-Rao manifold, the dual $\alpha$-connections, and the Jeffreys' prior) of the Pareto distributions has been studied in~\cite{abdel2003geometrical,sun2020bayesian,FisherRaoPareto-2022}.
The biparametric family of Pareto distributions $\{p_{s,a}(x)=(s-1) \frac{a^{s-1}}{x^{s}}, s>1, a>0\}$ equipped with the Fisher information metric yields a manifold of positive curvature~\cite{abdel2003geometrical,peng2007geometric} (and thus this contrasts with the manifolds of location-scale families which are always of non-positive curvature).

The Sharma-Mittal divergences~\cite{sharma1975new} between two densities $p$ and $q$ is a biparametric family of relative entropies is defined by
$$
D_{\alpha, \beta}[p:q]=\frac{1}{\beta-1}\left(\left(\int p(x)^{\alpha} q(x)^{1-\alpha} \mathrm{d} x\right)^{\frac{1-\beta}{1-\alpha}}-1\right), \forall \alpha>0, \alpha \neq 1, \beta \neq 1.
$$
The Sharma-Mittal divergence is induced from the Sharma-Mittal entropies which unifies the extensive R\'enyi entropies with the non-extensive Tsallis entropies~\cite{sharma1975new}.
The Sharma-Mittal divergences include the R\'enyi divergences ($\beta\rightarrow 1$) and the  Tsallis divergences ($\beta\rightarrow\alpha$), and in the limit case of $\alpha,\beta\rightarrow 1$ the Kullback-Leibler divergence~\cite{nielsen2011closed}. 
When both densities $p=p_{\theta_1}$ and $q=p_{\theta_2}$ belong to the same exponential family, we have the following closed-form formula~\cite{nielsen2011closed}:
$$
D_{\alpha, \beta}[p_{\theta_1}:p_{\theta_2}]=\frac{1}{\beta-1}\left(e^{-\frac{1-\beta}{1-\alpha} J_{F, \alpha}\left(\theta_1: \theta_2\right)}-1\right) .
$$

Thus we get the following theorem:

\begin{Theorem}
For $\alpha>0$, $\alpha\not=1$, $\beta\not=1$, the Sharma-Mittal divergence between two zeta distributions $p_{s_1}$ and $p_{s_2}$ is
$$
D_{\alpha,\beta}[p_{s_1}:p_{s_2}] = \frac{1}{\beta-1} \left(\left(\frac{\zeta(\alpha s_1+(1-\alpha)s_2)}{\zeta(s_1)^\alpha \zeta(s_2)^{1-\alpha}}\right)^{\frac{1-\beta}{1-\alpha}}-1\right).
$$
\end{Theorem}

\section{The Kullback-Leibler divergence between two zeta distributions}

It is well-known that the KLD between two probability mass functions of an exponential family amounts to a reverse Bregman divergence induced by the cumulant function~\cite{EFBD-2001}: $D_\KL[p_{s_1}:p_{s_2}]=B_F^*(\theta_1:\theta_2):=B_F(\theta_2:\theta_1)$ (with $\theta_1=s_1$ and $\theta_2=s_2$).
Furthermore, this Bregman divergence $B_F$ amounts to a Fenchel-Young divergence $Y_{F,F^*}$~\cite{nielsen2021geodesic} so that we have
$$
D_\KL[p_{s_1}:p_{s_2}]=B_F(\theta_2:\theta_1)=F(\theta(s_2))+F^*(\eta(s_1))-\theta(s_2)\eta(s_1):=Y_{F,F^*}(\theta(s_2):\eta(s_1)),
$$
where $F^*(\eta)$ denotes the Legendre convex conjugate of $F$, $\theta(s)=s$ and $\eta(s)=F'(\theta(s))=E_{p_s}[t(x)]=-E_{p_s}[\log x]$, see~\cite{EF-2014}.
Moreover, the convex conjugate $F^*(\eta(s))$ corresponds to the negentropy~\cite{EF-2010}: $F^*(\eta(s))=-H[p_s]$,
where the entropy of a zeta distribution $p_s$ is defined by:
$$
H[p_s]:=\sum_{i=1}^\infty p_s(i)\log\frac{1}{p_s(i)}.
$$

Using the fact that $\sum_{i=1}^\infty p_s(i)=1=\sum_{i=1}^\infty \frac{1}{i^s \zeta(s)}$, we can express the entropy as follows:
\begin{eqnarray*}
H[p_s] &=&\sum_{i=1}^\infty  \frac{1}{i^s\zeta(s)}\log i^s+\log(\zeta(s))\sum_{i=1}^\infty \frac{1}{i^s \zeta(s)},\\
&=& \sum_{i=1}^\infty  \frac{1}{i^s\zeta(s)}\log(i^s\zeta(s)).
\end{eqnarray*}
 
Since $F(\theta)=\log \zeta(\theta)$, we have $\eta(\theta)=F'(\theta)=\frac{\zeta'(\theta)}{\zeta(\theta)}$.
The function $\frac{\zeta'(\theta)}{\zeta(\theta)}$ has been tabulated in~\cite{walther1926anschauliches} (page 400).
Notice that the maximum likelihood estimator~\cite{barndorff2014information,bauke2007parameter} (MLE) of $n$ identically and independently (iid.) observations $x_1,\ldots, x_n$ is 
$$
\max_s \frac{1}{n}\sum_{i=1}^n \log p_s(x_i)  = \max_s -\frac{1}{n}\sum_{i=1}^n \log x_i-\log\zeta(s).
$$
Thus we get~\cite{seal1952maximum}.
$$
\hat{\eta}=\frac{\zeta'(\hat{\theta})}{\zeta(\hat{\theta})}=-\frac{1}{n}\sum_{i=1}^n \log x_i.
$$
See~\cite{deluca2013fitting} for the MLE of the truncated Pareto distributions.

\def\Var{\mathrm{Var}}
\begin{remark}
The Cram\'er-Rao lower bound (CRLB) states that the variance of any unbiased estimator $\hat s$ is greater or equal than the inverse of the Fisher information. Let $X=(X_1,\ldots,X_n)\sim_{\mathrm{iid}} \mathrm{Zeta}(s)$. Then $I_X=nI(s)$ and we have the CRLB:
$$
\Var[\hat{s}]\geq \frac{1}{n}I^{-1}(s)=\frac{\zeta^2(s)}{n (\zeta(s)\zeta''(s)-\zeta'(s)^2)}. 
$$ 
This is in accordance with Section~2 of~\cite{QuadraticZeta-1995}.
Moreover, the unbiased MLE matches exactly this bound only when dealing with exponential families~\cite{sundberg2019statistical}. 
 \end{remark}

The inverse of the zeta function $\zeta^{-1}(\cdot)$ has been studied in~\cite{kawalec2021inverse}.
An alternative estimator of the zeta parameter (called the quadratic distance estimator, QDE) has been proposed in~\cite{QuadraticZeta-1995}:
We consider the vector $X=\left(\log\frac{1}{2},\ldots,\log\frac{N-1}{N}\right)$ and the vector of log frequency ratio 
$Y=\left(\log\frac{f_2}{f_1},\ldots \log\frac{f_{N}}{f_{N-1}}\right)$ (where $f_i=\frac{n_i}{n}$ denotes the frequency of the $i$th item, the ratio of occurences of the $i$th item over the total number of items), and write the system of equations $Y=sX+(\epsilon_1,\ldots,\epsilon_{N-1})$. Thus the QDE amounts to a mere line fitting procedure. See also~\cite{corral2012practical}.

\begin{Proposition}[KLD between zeta distributions]
The Kullback-Leibler divergence between two zeta distributions can be written as:
\begin{eqnarray*}
D_\KL[p_{s_1}:p_{s_2}]&=&
\log(\zeta(s_2))-H[p_{s_1}]+s_2 E_{p_{s_1}}[\log x],\\
&=&  \log(\zeta(s_2)) -  \sum_{i=1}^\infty  \frac{1}{i^{s_1}\zeta(s_1)}\log(i^{s_1}\zeta(s_1)) - s_2 \frac{\zeta'(s_1)}{\zeta(s_1)}.
\end{eqnarray*}
\end{Proposition}

\begin{figure} 
\centering
\includegraphics[width=0.65\textwidth]{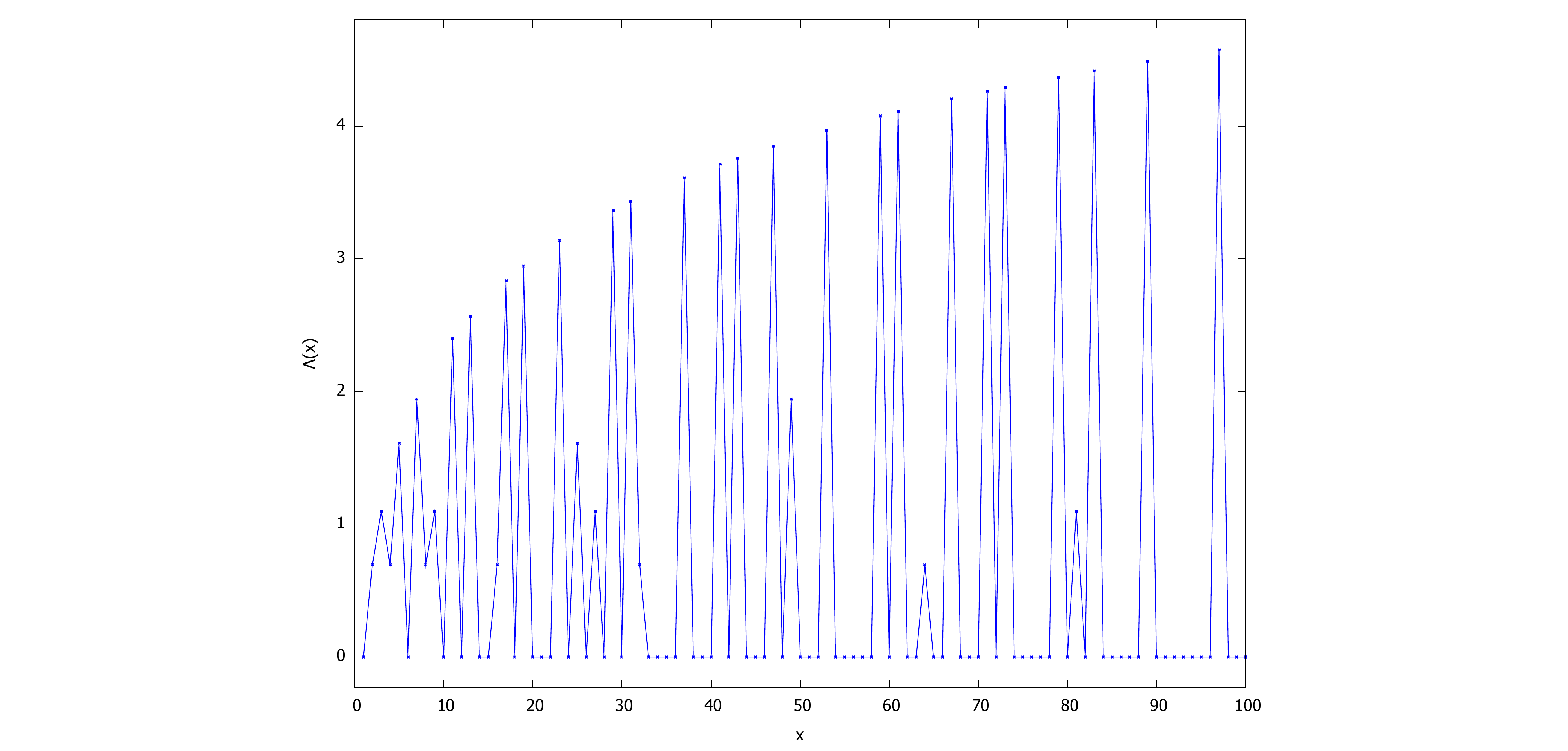}
\caption{Plot of the von Mangoldt function in the range $\{1,\ldots, 100\}$.}\label{fig:plotvonMangoldt}
\end{figure}

Moreover, the logarithmic derivative of the zeta function can be expressed using the von Mangoldt function~\cite{weisstein2002crc} (page 1850) for $\theta>1$:
$$
\eta(\theta)=\frac{\zeta'(\theta)}{\zeta(\theta)} = -\sum_{i=1}^{\infty} \frac{\Lambda(i)}{i^{\theta}},
$$
where $\Lambda(i)=\log p$ is $i=p^k$ for some prime $p$ and integer $k\geq 1$ and $0$ otherwise:
$$
\Lambda(i) = \begin{cases} \log p & \text{if }i=p^k \text{ for some prime } p \text{ and integer } k \ge 1, \\ 0 & \text{otherwise.} \end{cases}
$$
Figure~\ref{fig:plotvonMangoldt} displays a plot of the von Mangoldt function.
The von Mangoldt function satisfies the following identity:
$$
\log(n)=\sum_{i | n} \Lambda(i),
$$
where $i|n$ means $i$ divides $n$.

Notice that the zeta function can be calculated using Euler product formula: $\zeta(\theta)=\prod_{p:\mathrm{prime}} \frac{1}{1-p^{-\theta}}$.

\begin{Theorem}\label{thm:kldzeta}
The Kullback-Leibler divergence between two zeta distributions can be expressed using the real zeta function $\zeta$ and the von Mangoldt function $\Lambda$ as:
$$
D_\KL[p_{s_1}:p_{s_2}]=
 \log(\zeta(s_2)) -  \sum_{i=1}^\infty  \frac{1}{i^s\zeta(s)}\log(i^s\zeta(s)) + s_2 \sum_{i=1}^{\infty} \frac{\Lambda(i)}{i^{s_1}}.
$$
\end{Theorem}

\begin{Example}
Consider $s_1=4$ and $s_2=12$. 
Letting $\epsilon=0.9999$ and using Corollary~\ref{cor:klzetaapprox}, we get 
$$
D_\KL[p_{s_1}:p_{s_2}]\simeq D_{1-\epsilon}[p_{s_1}:p_{s_2}] = 0.430479743738878\ldots
$$

Let us now calculate the KLD using Theorem~\ref{thm:kldzeta}, we get
$\log(\zeta(s_2))=\log\frac{691\pi^12}{638512875}$, $H[p_{s_1}]\simeq 0.3337829096182664\ldots$ (using $100$ terms), and 
$\eta(s_1)=-0.06366938697034288\ldots$ (using $100$ terms) so that we have
\begin{eqnarray}
D_\KL[p_{s_1}:p_{s_2}] &=&
 \log(\zeta(s_2)) -  \sum_{i=1}^\infty  \frac{1}{i^s\zeta(s)}\log(i^s\zeta(s)) + s_2 \sum_{i=1}^{\infty} \frac{\Lambda(i)}{i^{s_1}},\\
&\simeq & 0.430495790304827\ldots
\end{eqnarray}
\end{Example}

It is well-known that the KLD between two arbitrarily close zeta distributions $p_s$ and $p_{s+\ds}$ amounts to half of the quadratic distance induced by the Fisher information:
$$
D_\KL[p_{s}:p_{s+\ds}]\approx \frac{1}{2} I(s)\ds^2,
$$
where
$$
I(s)=E_{p_s}[{(\log p_s(x))'}^2]=-E_{p_s}[(\log p_s(x))''],
$$
where the first-order and second-order derivatives are taken with respect to the parameter $s$.
Thus for uniorder exponential families, the Fisher information matrix is
$$
I(s)=-E_{p_s}[(\log p_s(x))'']=(\log\zeta(s))''=\frac{\zeta(s)\zeta''(s)-\zeta'(s)^2}{\zeta^2(s)}.
$$ 
This second-order derivative $(\log\zeta(s))''$ has been studied in~\cite{ZetaFI-2016}: Near $s=1$, we have 
$$
(\log\zeta(s))''=\frac{1}{(s-1)^2}+O(1),
$$ 
and this coincides with the FIM of the Pareto distribution (see Table~\ref{tab:comparison}).
We have
$$
I(s)=\sum_{n=1}^\infty \Lambda(n)\log(n)n^{-s}
$$
where $\Lambda(n)$ is the von Mangoldt  function.

\section{Comparison of the Zeta family with a Pareto subfamily}\label{sec:ZetaPareto}

The zeta distribution is also called the ``pure power-law distribution'' is in the literature~\cite{goldstein2004problems}.

We can compute the $\alpha$-divergences between two Pareto distributions $q_{s_1}$ and $q_{s_2}$ with fixed scale $1$ and respective shapes $s_1-1$ and $s_2-1$. In our case, the Pareto density writes $q_{s}(x)=\frac{s-1}{x^{s}}$ for $x\in\calX=(1,\infty)$.
The family of such Pareto distributions forms a continuous exponential family with natural parameter $\theta=s$, sufficient statistic $t(x)=-\log(x)$, and convex cumulant function $F(\theta)=-\log(\theta-1)$ for $\theta\in\Theta=(1,\infty)$.
Thus we have~\cite{BR-2011}: 
\begin{eqnarray*}
I_\alpha[q_1:q_2]=\int q_{s_1}(x)^{\alpha} q_{s_2}(x)^{1-\alpha}\dx &=& \exp(-J_{F,\alpha}(\theta_1:\theta_2)),\\
 &=& \frac{\alpha s_1+(1-\alpha)s_2}{s_1^\alpha s_2^{1-\alpha}},
\end{eqnarray*}
and we get the following closed-form for the $\alpha$-divergences between two Pareto distributions $q_{s_1}$ and  $q_{s_2}$:
$$
D_\alpha[q_{s_1}:q_{s_2}] = \frac{1}{\alpha(1-\alpha)} \left(1-\frac{\alpha s_1+(1-\alpha)s_2}{s_1^\alpha s_2^{1-\alpha}}\right).
$$
The moment parameter is $\eta(\theta)=F'(\theta)=-\frac{1}{\theta-1}$ so that $\theta(\eta)=1-\frac{1}{\eta}$ and 
$F^*(\eta)=\theta(\eta)\eta-F(\theta(\eta))=\eta-1-\log(-\eta)$.
It follows that the KLD is
$$
D_\KL[q_{s_1}:q_{s_2}]=B_F(\theta_2:\theta_1)=\log\left(\frac{s_1-1}{s_2-1}\right)+\frac{s_2-s_1}{s_1-1}.
$$
The differential entropy of the Pareto distribution $q_s$ is 
$$
h[q_s]=-\int_1^\infty q_s(x)\log q_s(x)\dx=-F^*(\eta(s))
$$ 
with $\eta(s)=-\frac{1}{s-1}$. We find that 
$$
h[q_s]=1+\frac{1}{s-1}-\log(s-1).
$$

The Pareto distributions form a discrete exponential family and are thus maximum entropy distributions under the moment constraints 
$-E[\log x]=\eta$:

$$
\max \left\{h(q) \quad :\quad E[\log x]=-\eta \right\}.
$$
The differences with the Zeta distribution is that the support is considered $(1,\infty)$ instead of $\bbN$ and that the entropy is the differential entropy instead of the discrete entropy. Notice that the differential entropy may be negative (e.g., for the Pareto distributions when $s$ is large) but never the discrete entropy.

\begin{Example}
For comparison, we calculate the KLD between two Pareto distributions with parameters $s_1=4$ and $s_2=12$.
We find 
$$
D_\KL[q_{s_1}:q_{s_2}]=\log\frac{3}{11}+\frac{8}{3} \simeq 1.367383682536406\ldots
$$
\end{Example}

Table~\ref{tab:comparison} compares the discrete exponential family of zeta distributions with 
the continuous exponential family of Pareto distributions with fixed scale $1$.

Since the information-theoretic distances between zeta distributions are computationally demanding, one can also investigate both fast(er) lower and upper bounds on these distances. Bounds on the $f$-divergences (including the $\alpha$-divergences) between Zipf-Mandelbrot distributions have been studied in~\cite{ZipfMandelbrotFDiv-2018,adil2020new}.

In general, it is interesting to consider discrete counterparts of continuous exponential families.
For example, the discrete Gaussian distributions or discrete normal distributions defined as maximum entropy distributions have been studied in~\cite{DiscreteGaussianThetaSiegel-2019,LatticeGaussian-2022}.
The log-normalizer or cumulant function of the discrete Gaussian distributions are related to the Riemann theta function~\cite{deconinck2004computing}.
Given a prescribed sufficient statistics $t(x)$, we may define the continuous exponential family wrt the Lebesgue measure $\mu$ as the probability density functions $p(x)$ maximizing the differential entropy under the moment constraint $E_p[t(x)]=\eta$. 
The corresponding discrete exponential family is obtained by the distributions with probability mass functions maximizing Shannon entropy under the moment constraint $E_p[t(x)]=\eta$.
Notice that the raw (uncentered) moments $\mu_k$ of the zeta distributions are
$$
\mu_k = E[X^k] = \frac{1}{\zeta(s)}\sum_{i=1}^\infty \frac{1}{i^{s-k}}
=\left\{
\begin{matrix}
\zeta(s-k)/\zeta(s) & \textrm{for}~k < s-1 \\
\infty & \textrm{for}~k \ge s-1
\end{matrix}
\right.
$$

\section{Clustering finite sets of Zipf's distributions}\label{sec:clustering}

Consider a finite set $\calP=\{p_{\theta_1,N_1}, \ldots, p_{\theta_n,N_n}\}$ of $n$ Zipf's distributions with corresponding discrete supports $\calX_1=\{1,\ldots, N_1\},\ldots,\calX_n=\{1,\ldots,N_n\}$.
For example, to fix ideas, we may consider the set of $100$ Zipf's distributions obtained by analyzing the word frequency of translations of the Holy bible (see Table~1 of~\cite{ZipfHolyBible-2017} with a short excerpt displayed in Table~\ref{tab:hbl}). Each translation in a natural language uses a vocabulary of $N$ distinct words and is modeled by a Zipf distribution $p_{\theta,N}$.
These Zipf's distributions somehow characterize some intrinsic properties of natural languages~\cite{ferrer2005variation}, and we may cluster these Zipf's distributions to interpret how corresponding languages are similar~\cite{gamallo2020measuring} or not.
We may cluster either using the agglomerative hierarchical clustering or partition-based $k$-means or $k$-centers algorithms~\cite{nielsen2016introduction}.

\begin{table}
\caption{Some parameters $\theta>0$ of Zipf's distributions obtained as word ranking-frequency distributions of the Holy Bible translations. Data excerpt extracted from Table~1 of~\cite{ZipfHolyBible-2017}.}\label{tab:hbl}
\centering
\vskip 0.5cm
\begin{tabular}{lll}
Natural language & $\theta$ & N \\ \hline
English & 1.258 & 12702\\
French & 1.161 & 24716\\
Japanese & 0.774 & 30785 \\
Danish & 1.158 & 26290 \\
Chinese & 0.792 & 1699\\
Finnish & 0.997 & 54863\\
... & ... & ...
\end{tabular}
\end{table}

\begin{figure} 
\centering
\includegraphics[width=0.45\textwidth]{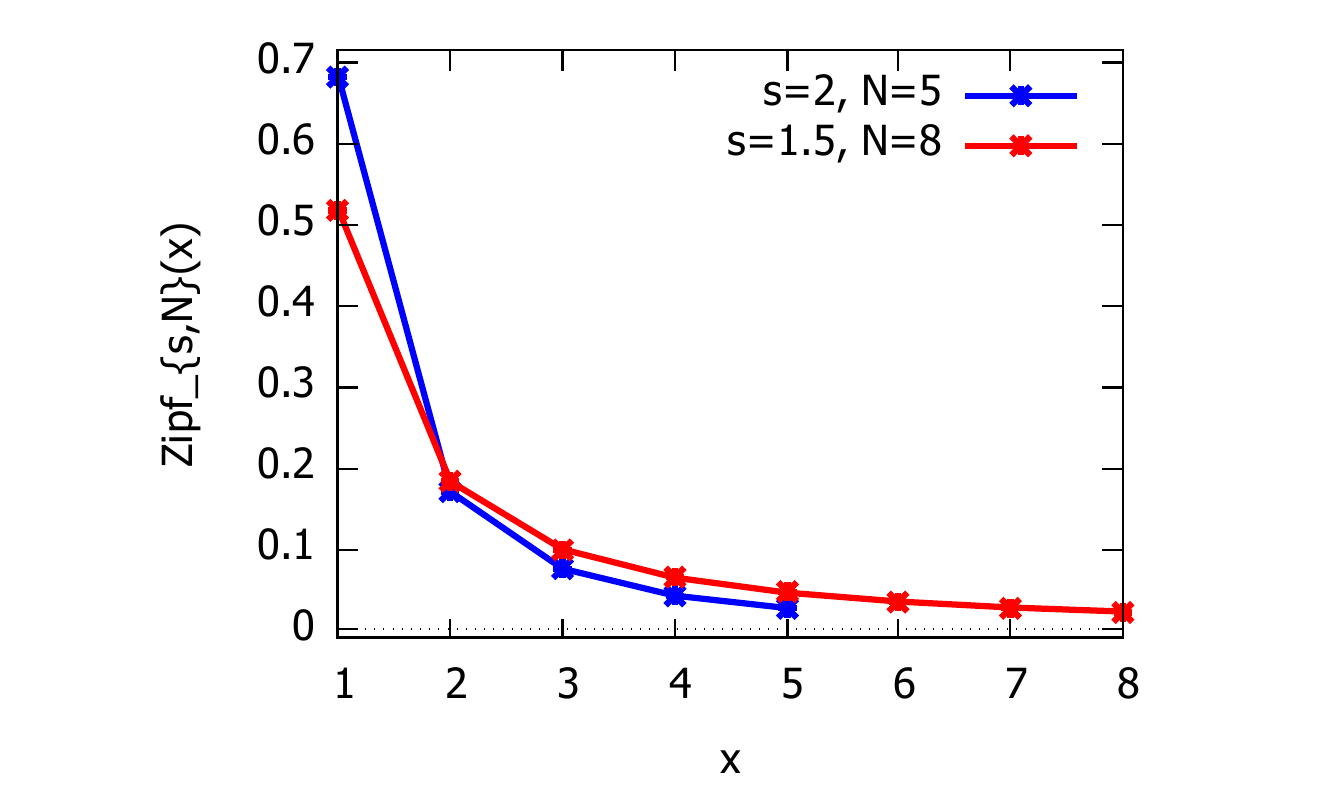}
\caption{Two probability mass functions of Zipf's distribution with different supports.}\label{fig:plotZipf}
\end{figure}

Consider the $k$-means algorithm which partitions $\calP=\uplus_{i=1}^k \calP_i$ into $k$ pairwise disjoint subsets, with each subset $\calP_j$ summarized by a zeta prototype distribution $q_{s_j}$ with full support $\bbN$. 
Lloyd's heuristic of $k$-means consists in iteratively associating to each Zipf's distribution $p_{\theta_i,N_i}$ its closest zeta distribution $q_{s_j}$ with respect to the Kullback-Leibler divergence, and then update the cluster zeta distribution prototypes by taking their cluster centroids with respect to the  Kullback-Leibler divergence. 
This clustering algorithm yields is an extension of the Bregman $k$-means algorithm~\cite{banerjee2005clustering} once the KLD between a Zipf's distribution $p_{\theta,N}$ (i.e., a truncated zeta distribution) and a zeta distribution $q_s$ is identified to a duo Bregman divergence~\cite{duodivergence-2022}:
\begin{equation}\label{eq:KLDZipfZeta}
D_\KL[p_{\theta,N}:q_s]=B_{F_2,F_1}(s:\theta)=F_2(s)-F_1(\theta)-(s-\theta)F_1'(\theta),
\end{equation}
where $F_2(s)=\log \zeta(s)$, $F_1(\theta)=\log H_{N,\theta}$ where 
$$
H_{N,\theta}=\sum_{i=1}^N \frac{1}{i^\theta}
$$
denotes the generalized harmonic number with
$$
F_1'(\theta)=\frac{H_{N,\theta}'}{H_{N,\theta}}=-\sum_{i=1}^N \frac{\log i}{i^\theta\, H_{N,\theta}}.
$$
Figure~\ref{fig:plotZipf} displays two PMFs of Zipf's distributions with different supports.
The set of Zipf's distribution with fixed support $\{1,\ldots, N\}$ form an exponential family $\calE_N$, and thus the set of all Zipf exponential families $\cup_{N=1}^{\infty} \calE_N$ form a ``foliated  exponential family'' (like the set of Weibull distributions with shape parameter $k$ in $\{1,\ldots\}$ form another foliated exponential family) with the exponential family of zeta distributions in the limit case $N\rightarrow\infty$.

Thus the KLD between a Zipf distribution $p_{\theta,N}$ and a zeta distribution $q_s$ can be calculated in closed-form using Eq.~\ref{eq:KLDZipfZeta}:
\begin{equation}
D_\KL[p_{\theta,N}:q_s]=\log \frac{\zeta(s)}{\sum_{i=1}^N \frac{1}{i^\theta}}+(s-\theta)\sum_{i=1}^N \frac{\log i}{i^\theta\, H_{N,\theta}}.
\end{equation}

\begin{remark}
More generally, we may consider two truncated zeta distributions $q_{s_1}^{a_1,b_1}(x)=\frac{1}{x^{s_1} \zeta^{a_1,b_1}(s_1)}$ and 
$q_{s_1}^{a_2,b_2}=\frac{1}{x^{s_1} \zeta^{a_2,b_2}(s_2)}$ with $\zeta^{a,b}(s)=\sum_{i=a}^b \frac{1}{x^s}$.
When $[a_1,b_1]\subseteq [a_2,b_2]$, we may apply formula  Eq.~\ref{eq:KLDZipfZeta} with $F_1(s)=\log \zeta^{a_1,b_1}(s)$ and 
$F_2(s)=\log \zeta^{a_2,b_2}(s)$.
\end{remark}

Lloyd's heuristic minimizes the following energy:
$$
\sum_{i=1}^n \min_{j\in\{1,\ldots,k\}} D_\KL[p_{\theta_i}:s_j] = \sum_{i=1}^n \min_{j\in\{1,\ldots,k\}} B_{F_2,F_1}(s_j:\theta_i).
$$

To calculate the cluster prototype $p_{s_i}$ corresponding to cluster $\calP_i$, we need to solve the following generic optimization problem:
$$
\min_{s\in(0,\infty)} \sum_{p_{\theta,N}\in\calP_i} D_\KL[p_{\theta,N}:q_s] = \min_{s\in(0,\infty)} \sum_{p_{\theta,N}\in\calP_i} B_{F_2,F_1}(s:\theta).
$$
We get $\eta_s:=\frac{\zeta'(s)}{\zeta(s)}= \frac{1}{|\calP_i|} \sum_{p_{N,\theta}\in\calP_i} \frac{H_{N,\theta}'}{H_{N,\theta}}$.
Notice that we can also cluster just by a careful initialization using $k$-means++~\cite{vassilvitskii2006k} extended to any arbitrary divergence~\cite{nielsen2015total} (here a duo Bregman divergence).
However, although the prototype distributions $q_{s_i}$ are parameterized by a single parameter $s_i$, we cannot use 
 the optimal interval clustering relying on dynamic programming~\cite{nielsen2014optimal} since the supports $\calX_1, \ldots,\calX_n$ may be different, and the clusters in the optimal $k$-means may have disjoint intervals.

\begin{remark}
When all Zipf's distributions have coinciding support $\calX$, we get the interval clustering property of $k$-means (since in that case, Bregman Voronoi diagrams have connected cells) and may use the optimal dynamic programming algorithm~\cite{nielsen2014optimal}.
\end{remark}

We may also consider the Bhattacharyya distance between two Zipf's distributions $p_{\theta_i,N_i}$ and $p_{\theta_j,N_j}$ by considering the common support $\{1,\ldots,\min\{N_i,N_j\}\}$:
$$
D_\Bhat[p_{\theta_i,N_i}:p_{\theta_i,N_i}]=-\log \sum_{x=1}^{\min\{N_i,N_j\}} \sqrt{p_{\theta_i,N_i}(x)\, p_{\theta_j,N_j}(x)}.
$$
We get a duo Jensen divergence (see Theorem~2 of~\cite{duodivergence-2022}).

Yet another approach is to convert the Zipf's distributions $p_{N_i,s_i}$ into Zeta distributions $p_{s_i}$.
This conversion is motivated by Seal~\cite{seal1947probability}'s estimator of the parameter of a Zeta distribution from an iid random sample of size $n$:
$$
\hat{s}_{\mathrm{Seal}}=\frac{\log\frac{f_1}{f_2}}{\log 2},
$$
where $f_i=\frac{n_i}{n}$ denotes the $i$-th highest frequency.  Since $\frac{f_1}{f_2}=2^s$ for the zeta distribution, we get $\hat{s}_{\mathrm{Seal}}=s$.
This estimator has variance $\Var[\hat{s}_{\mathrm{Seal}}]=\frac{\zeta(s)(1+2^s)}{n\log^2 2}$.

\bibliographystyle{plain}
\bibliography{ZetaDistributionsBIBv3}

\appendix

\section{Code snippets in {\sc Maxima}}\label{sec:maxima}

We give below some code snippets in {\sc Maxima} (\url{https://maxima.sourceforge.io/}) to numerically calculate the examples reported in this technical report.

\subsection*{Code snippet 1}
{\footnotesize
\begin{verbatim}
PMFzeta(x,s):=1.0/((x**s)*zeta(s));
s1:4;
s2:12;
alpha:1/2;

expminusJ(alpha,s1,s2):=zeta(alpha*s1+(1-alpha)*s2)/((zeta(s1)**alpha)*(zeta(s2)**(1-alpha)));
(1/(alpha*(1-alpha)))*(1-expminusJ(alpha,s1,s2));
bfloat(%);

(1/(alpha*(1-alpha)))*(1-sum((PMFzeta(x,s1)**alpha)*(PMFzeta(x,s2)**(1-alpha)), x, 1, 20));
bfloat(%);
\end{verbatim}
}

\subsection*{Code snippet 2}

{\footnotesize
\begin{verbatim}
/* von Mangoldt function */
Mangoldt(i):=if integerp(i)
  then block([fct],fct:ev(ifactors(i),factors_only:true),
                   if length(fct)=1 then log(fct[1]) else 0)
  else 'Mangoldt(i)

s1:4;
s2:12;
PMFzeta(x,s):=1.0/((x**s)*zeta(s));
s1:4;
s2:12;
alpha:0.99999;
/* Bhattacharyya coefficient */
expminusJ(alpha,s1,s2):=zeta(alpha*s1+(1-alpha)*s2)/((zeta(s1)**alpha) * (zeta(s2)**(1-alpha)));
/* alpha-divergence*/
(1/(alpha*(1-alpha)))*(1-expminusJ(alpha,s1,s2));
bfloat(%);

/* number of terms in the sums */
nbsum:100;

H(s):=sum((1/((i**s)*zeta(s))*log((i**s)*zeta(s))),i,1,nbsum);
eta(s):=sum(-float(Mangoldt(i))/(i**s),i,1,nbsum);
KL(s1,s2):= log(zeta(s2))-H(s1)-s2*eta(s1); 

H(s1);bfloat(%);
eta(s1); bfloat(%);
KL(s1,s2); bfloat(%);
\end{verbatim}
}

\subsection*{Code snippet 3}

\begin{verbatim}
s1:4;
s2:12;
pdfPareto(x,s):=(s-1)/(x**s);

integrate(pdfPareto(x,s1)*log(pdfPareto(x,s1)/pdfPareto(x,s2)),x,1,inf);
ratsimp(%);
bfloat(%);

KLParetoCF(s1,s2):=log((s2-1)/(s1-1))+(s1-s2)/(s2-1);
KLParetoCF(s2,s1);
bfloat(%);
\end{verbatim}

\subsection*{Code snippet 4}

Plotting zeta PMFs:

\begin{verbatim}
PMFzeta(x,s):=1.0/((x**s)*zeta(s));
xmax:5;
xx:makelist(x,x,1,xmax)$
yy101:makelist(PMFzeta(x,1.1),x,1,xmax)$
yy15:makelist(PMFzeta(x,1.5),x,1,xmax)$
yy2:makelist(PMFzeta(x,2),x,1,xmax)$
yy25:makelist(PMFzeta(x,2.5),x,1,xmax)$
yy3:makelist(PMFzeta(x,3),x,1,xmax)$
plot2d([[discrete,xx,yy101],[discrete,xx,yy15],[discrete,xx,yy2],
[discrete,xx,yy25],[discrete,xx,yy3]],[xlabel,"x"],
 [ylabel,"Zeta_s(x)"],[legend,"s=1.1", "s=1.5", "s=2","s=2.5","s=3"],
[style,  [linespoints,3,3],[linespoints,3,3],
 [linespoints,3,3], [linespoints, 3,3],[linespoints, 3,3]],
[point_type,asterisk]);
\end{verbatim}

\section{Drawing zeta variates and truncated Pareto variates}\label{sec:variate}

\subsection{Zeta variates}
We describe the acceptance/rejection method given in~\cite{rDEV86a,gentle2003random} to draw zeta variates:

\begin{itemize}
	\item Draw $u_1\sim\mathrm{Unif}(0,1)$ and $u_2\sim\mathrm{Unif}(0,1)$
	
	\item Let $x=\floor{u_1^{-\frac{1}{s-1}}}$ and $t=\left(1+\frac{1}{x}\right)^{s-1}$.
	
	\item Accept $x$ if $x\leq \frac{t}{t-1} \frac{2^{s-1}-1}{2^{s-1}u_2}$
\end{itemize}

\subsection{Truncated Pareto variates}
We consider a truncated Pareto distribution with support $\calX=(a,b)$ for $a<b$.
The probability density function of such a truncated Pareto distribution is
$$
q_s^{a,b}(x)=\frac{(s-1)a^{s-1}}{\left(1-\left(\frac{a}{b}\right)^{s-1}\right)\ x^{s}}.
$$

Using the inverse transform method from a uniform variate $u\sim\mathrm{Unif}(0,1)$, we get a truncated Pareto variate:
$$
x=ab \left(b^{s-1}-u(b^{s-1}-a^{s-1})\right)^{-\frac{1}{s-1}}.
$$

\end{document}